\documentclass[3p,times,twocolumn]{elsarticle}

\usepackage{ecrc}

\volume{00}

\firstpage{1}

\journalname{Subatomic Particles and Cosmology}

\runauth{Amit Pathak}
\runauth{}

\jid{spc}

\jnltitlelogo{Subatomic Particles and Cosmology}

\usepackage{amssymb}
\usepackage{amsmath}
\usepackage{lineno}
\usepackage{graphicx}
\usepackage{fancyhdr}

\usepackage[figuresright]{rotating}

\begin{document}

\begin{frontmatter}



\dochead{}

\title{Search for new physics with baryons at BESIII}


\author{Amit Pathak$^{1}$ on behalf of the BESIII Collaboration}


\address{$^{1}$Department of Physics, Chung-Ang University, Seoul, 06974, South Korea}


\begin{abstract}
The BESIII experiment at BEPCII provides a clean environment to test baryon–dark sector connections and baryon-number violation. Using $10^{10}$ $J/\psi$ events, searches were performed for $\Sigma^{+}\!\to\!p+\text{inv}$, $\Xi^{-}\!\to\!\pi^{-}+\text{inv}$, and $\bar{\Lambda}$–$\Lambda$ oscillations. No excess was observed, yielding $\mathcal{B}(\Sigma^{+}\!\to\!p+\text{inv})<3.2\times10^{-5}$, $\mathcal{B}(\Xi^{-}\!\to\!\pi^{-}+\text{inv})<(4$–$7)\times10^{-5}$, and $P(\Lambda)<4.4\times10^{-6}$ (90\% C.L.). These are the most stringent limits on invisible baryon decays and $\Delta B=2$ transitions in strange baryons, probing new physics beyond the Standard Model.
\end{abstract}

\begin{keyword}
BESIII, J/$\psi$, $\Sigma$, $\Xi$, $\Lambda$, Baryons, New Physics, Dark Matter

\end{keyword}

\end{frontmatter}


\section{Introduction}
\label{sec1}
The Standard Model (SM) of particle physics provides an exceptionally successful description of the known elementary particles and their interactions. Precision tests at colliders and low-energy experiments have validated its predictions over
a wide range of energy scales. Nevertheless, the SM is incomplete. It does not explain the existence of dark matter \cite{Trimble1987, Bergstrom2000, Zasov2017, Bertone2005}, the observed baryon asymmetry of the Universe \cite{deVries2024, ChoeJo2024, Vien2025}, or the origin of neutrino masses and mixing \cite{PDG2025nu, Chakraborty2024, Vien2025nu}. These open problems strongly motivate searches for new physics beyond the SM.

Among these questions, the nature of dark matter and its possible connection to ordinary baryonic matter is particularly intriguing. Cosmological observations
indicate that the dark matter density exceeds the baryonic matter density by a factor of about five,
\begin{equation}
\rho_{\rm DM} \simeq 5.4\,\rho_{\rm baryon},
\end{equation}
suggesting a potential common origin of visible and dark matter
\cite{Planck2020}. This numerical coincidence has inspired a broad class of models in which the dark sector carries a baryon-like quantum number, conserved through a new gauge symmetry \cite{Zurek2014,Essig2013}. Such scenarios predict
weak interactions between SM baryons and dark states, leading naturally to rare or invisible baryon decays.

Baryons therefore constitute a powerful and complementary probe of physics beyond the SM. In particular, hyperon decays are sensitive to flavor-changing neutral current processes, invisible final states, and baryon-number-violating interactions. Within the SM, these processes are either forbidden or extremely suppressed by the Glashow--Iliopoulos--Maiani (GIM) mechanism, with branching fractions far below current experimental sensitivity
\cite{He2019}. Any observable signal in these channels would thus provide a clear indication of new physics.

High-intensity electron--positron colliders offer an ideal environment for such searches. The BESIII experiment at the BEPCII collider has collected unprecedented samples of charmonium decays, including more than $10^{10}$
$J/\psi$ events \cite{BESIII_Dataset}. These data enable clean production of baryon-antibaryon pairs and precision studies of rare baryon decays using recoil
and double-tag techniques.

Motivated by these opportunities, the BESIII collaboration has performed a series of searches for new physics in the baryon sector, including searches for invisible hyperon decays \cite{SigmaInvisible,XiInvisible} and baryon-number--violating processes such as $\bar{\Lambda}$--$\Lambda$
oscillations \cite{LambdaOsc}. The resulting constraints provide important and complementary limits on models involving dark baryons, axion-like particles, dark photons, and neutrino-portal interactions.
\section{BESIII Results: Searches for New Physics with Baryons}

\subsection{\textbf{Search for a Massless Particle in $\Sigma^{+}\to p + \mathrm{invisible}$}}

In the SM of particle physics, the decay $\Sigma^{+}\to p\nu\bar{\nu}$ is a flavor-changing neutral current (FCNC) process occurring only at the loop level and is strongly suppressed by the Glashow--Iliopoulos--Maiani mechanism.
The predicted branching fraction of such decays is below $10^{-11}$ \cite{Tandean2019}, rendering the decay experimentally inaccessible. New light particles beyond the SM, such as axion-like particles or dark photons, can significantly enhance the decay rate,
making $\Sigma^{+}\to p + \mathrm{invisible}$ a sensitive probe of new physics.

This search is performed using $(1.0087 \pm 0.0044)\times10^{10}$ $J/\psi$ events collected with the BESIII detector. The decay $J/\psi\to\Sigma^{+}\Sigma^{-}$ is
reconstructed using a double-tag technique, where the $\Sigma^{-}$ is fully reconstructed in the mode $\Sigma^{-}\to\bar{p}\pi^{0}$ and the signal decay
$\Sigma^{+}\to p + \mathrm{invisible}$ is identified via the recoil system. The signal signature consists of a single proton and vanishing extra energy in the electromagnetic calorimeter.

\begin{figure}[t]
  \centering
  \includegraphics[width=0.45\textwidth]{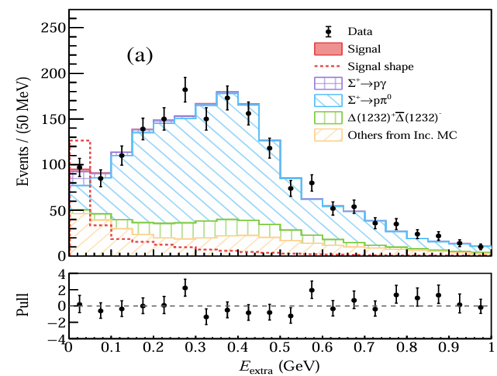}
  \includegraphics[width=0.47\textwidth]{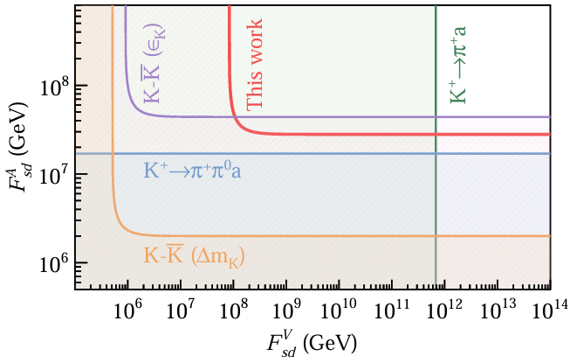}
  \caption{Upper: Fit to the extra EMC energy distribution used to extract the $\Sigma^{+}\to p + \mathrm{invisible}$ signal. Below: Constraints on effective FCNC couplings derived from the measured upper limit.}
  \label{fig:sigma_inv}
\end{figure}

No significant excess over background is observed as can be seen in Figure~\ref{fig:sigma_inv} (upper plot). The red dahsed line shows that invisible signal peaks at zero and the observed measured value $(0.6\pm1.5)\times10^{-5}$. Hence, an upper limit at 90\% confidence level is set:
\[
\mathcal{B}(\Sigma^{+}\to p + \mathrm{invisible}) < 3.2\times10^{-5}.
\]
This result constitutes the first FCNC search with missing energy in the baryon sector and provides competitive constraints on axion--fermion and dark mediator couplings (Figure-\ref{fig:sigma_inv} (lower plot)).

\subsection{\textbf{Search for Dark Baryons in $\Xi^{-}\to\pi^{-} + \mathrm{invisible}$}}

The observed coincidence between dark matter and baryon energy densities, $\rho_{\mathrm{DM}}\simeq5.4\,\rho_{\mathrm{baryon}}$, has motivated a broad class of models in which dark matter carries a conserved baryon number or baryon-like quantum number. In such scenarios, baryon number is shared between the visible and dark sectors, allowing SM baryons to decay into final states
containing dark baryons while conserving the total baryon number. These models predict hyperon decays with invisible final states and offer a natural connection between baryonic matter and dark matter.

The decay $\Xi^{-}\to\pi^{-}+\chi$, where $\chi$ denotes a dark baryon, provides a clean experimental probe of such scenarios. In the SM, no decay with this topology and an invisible massive particle exists, and therefore any observable signal would constitute a clear indication of physics beyond the SM. Compared to nucleon invisible decay searches, hyperon decays offer complementary sensitivity due to different quark-level transitions and hadronic matrix elements.

This analysis is based on $(10.087 \pm 0.044)\times10^{9}$ $J/\psi$ events collected with the BESIII detector at a center-of-mass energy of $\sqrt{s}=3.097$~GeV. Hyperon pairs are produced via $J/\psi\to\Xi^{-}\Xi^{+}$. The $\Xi^{+}$ is fully reconstructed using a double-tag technique in the decay chain $\Xi^{+}\to\pi^{+}\Lambda$, followed by $\Lambda\to p\pi^{-}$. The signal decay $\Xi^{-}\to\pi^{-}+\chi$ is identified using a recoil-mass technique, which allows the presence of an invisible
particle to be inferred without direct detection.

\begin{figure}[t]
  \centering
  \includegraphics[width=0.49\textwidth]{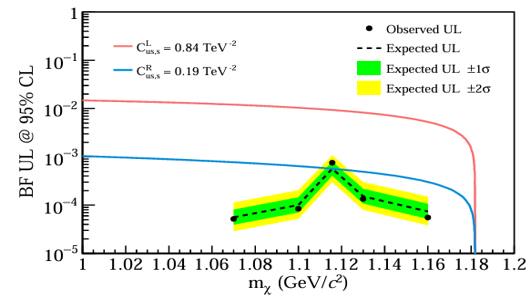}
  \includegraphics[width=0.48\textwidth]{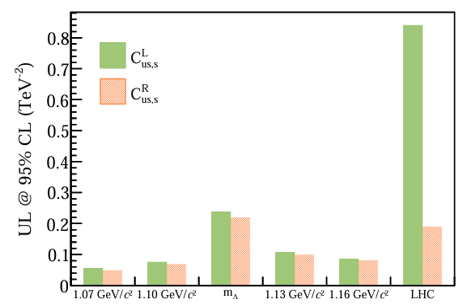}
  \caption{Upper: Diagram of the decay $\Xi^{-}\to\pi^{-}+\chi$, where $\chi$ is a dark baryon. Below: Upper limits on the branching fraction as a function of the assumed dark-baryon mass.}
  \label{fig:xi_inv}
\end{figure}

The search is performed for dark-baryon masses in the range
$1.07$--$1.16$~GeV/$c^{2}$ (Figure~\ref{fig:xi_inv} ), corresponding to kinematically allowed values in $\Xi^{-}$ decays. For each assumed dark-baryon mass, signal efficiencies and background contributions are evaluated, and the recoil mass spectrum is examined for deviations from the background-only expectation. No statistically significant signal is observed for any mass hypothesis.

Upper limits at the 90\% confidence level are set on the branching fraction $\mathcal{B}(\Xi^{-}\to\pi^{-}+\chi)$ as a function of the dark-baryon mass. The limits are found to be at the level of $10^{-5}$ across the entire scanned mass range, representing the most stringent constraints to date on dark-baryon production in hyperon decays. These results significantly restrict models of
baryonic dark matter and provide complementary constraints to those obtained from invisible nucleon decay searches and high-energy collider experiments.

This study demonstrates the strong potential of hyperon decays at
high-intensity $e^{+}e^{-}$ colliders to probe baryonic dark sectors and highlights the unique role of the BESIII experiment in exploring new physics through rare baryon processes.

\subsection{\textbf{Search for $\bar{\Lambda}$--$\Lambda$ Oscillations}}

The violation of baryon number is a fundamental ingredient for explaining the
matter–antimatter asymmetry of the Universe, as outlined by Sakharov’s
conditions. While neutron–antineutron oscillations have long been considered a
classic portal to $\Delta B=2$ interactions, oscillations between a baryon and
its antibaryon have not been experimentally explored until recently. In
particular, $\bar{\Lambda}$--$\Lambda$ oscillations provide a novel probe of
$\Delta B=2$ processes involving strange quarks, complementary to nucleonic
searches.

The first experimental search for $\bar{\Lambda}$--$\Lambda$ oscillations was
reported by the BESIII Collaboration using $1.31\times10^{9}$ $J/\psi$ events
collected at $\sqrt{s}=3.097$~GeV. In this analysis, coherent
$\bar{\Lambda}\Lambda$ pairs are produced in the decay
$J/\psi\to pK^{-}\bar{\Lambda}+$c.c. as the right-sign (RS) sample, while
wrong-sign (WS) final states such as $J/\psi\to pK^{-}\Lambda$ arise only if
$\bar{\Lambda}$ oscillates to $\Lambda$ before decay. The time evolution of
$\bar{\Lambda}$--$\Lambda$ oscillations can be described by a Schrödinger-like
formalism with an off-diagonal transition mass $\delta m_{\Lambda\bar{\Lambda}}$,
and the time-integrated oscillation probability $P(\Lambda)$ is related to
$\delta m_{\Lambda\bar{\Lambda}}$ and the $\Lambda$ lifetime $\tau_{\Lambda}$
via
\[
\delta m_{\Lambda\bar{\Lambda}} \simeq \sqrt{\frac{P(\Lambda)}{2\,\tau_{\Lambda}^{2}}}\,.
\]

No significant WS signal is observed after full event selection, and the upper limit on the time-integrated oscillation probability is determined to be $P(\Lambda)<4.4\times10^{-6}$ at 90\% confidence level, corresponding to $\delta m_{\Lambda\bar{\Lambda}}<3.8\times10^{-18}$~GeV \cite{LambdaOsc}. This measurement established the first experimental constraint on $\bar{\Lambda}$–$\Lambda$ oscillations and opened a new class of searches for
baryon-number-violating processes in the strange baryon sector.

More recently, BESIII has improved these limits using a larger data sample of $(10087\pm44)\times10^{6}$ $J/\psi$ events. Using the same coherent production approach in the decay $J/\psi\to\Lambda\bar{\Lambda}$, the updated study sets a more stringent upper limit on the integrated oscillation probability, $P(\Lambda)<1.4\times10^{-6}$, and on the transition mass, $\delta m_{\Lambda\bar{\Lambda}}<2.1\times10^{-18}$~GeV, at 90\% confidence level. This corresponds to an oscillation time limit of $\tau_{\rm osc}>3.1\times10^{-7}$\,s.

The improved limits represent a factor of roughly three reduction in
$\delta m_{\Lambda\bar{\Lambda}}$ compared to the previous measurement \cite{PRD111052014}, and advance the sensitivity to $\Delta B=2$ processes involving strange baryons by directly probing oscillation dynamics rather than nucleon decay alone. These constraints are still many orders of magnitude weaker than those on neutron–antineutron oscillations (for which the oscillation time limit is
$\gtrsim8.6\times10^{7}\,$s), but they provide the first direct experimental bounds on baryon-number violation in the hyperon sector and complement nucleonic searches and cosmological constraints on baryon-number-violating interactions. This body of work establishes hyperon oscillations as a viable search channel for future intensity-frontier facilities.

\begin{figure}[ht]
  \centering
  \includegraphics[width=0.45\textwidth]{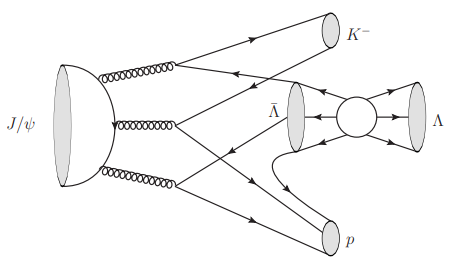}
  \includegraphics[width=0.45\textwidth]{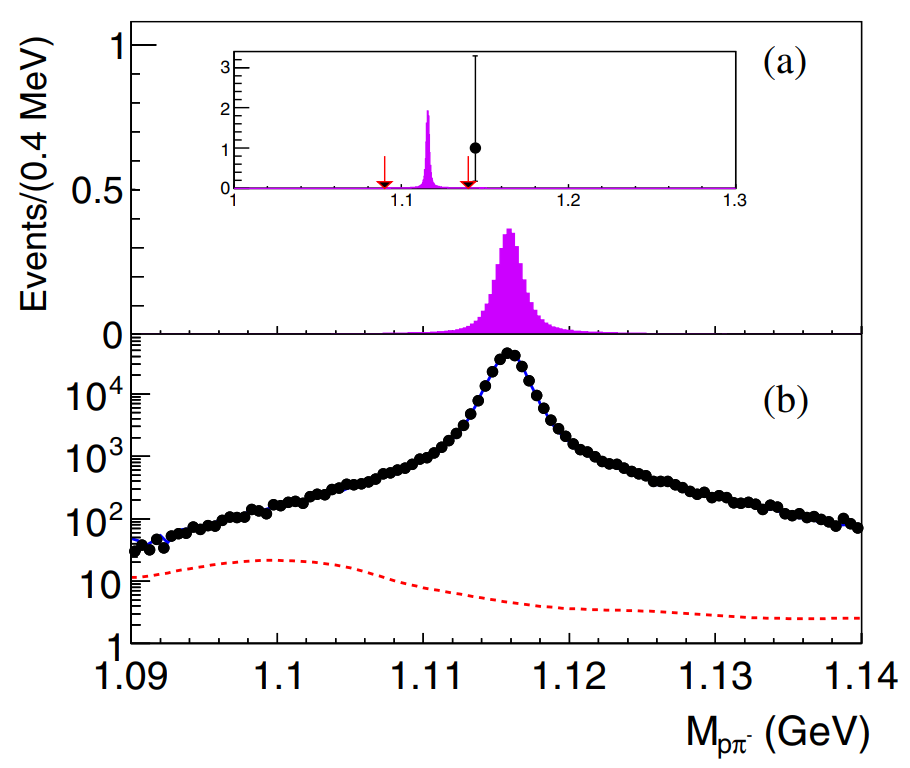}
  \caption{Upper: Illustration of $\bar{\Lambda}$--$\Lambda$ oscillation. Below: Invariant mass distribution of $M_{p\pi^{-}}$ for (a) WS events in the signal region and (insert) over the full span, where the filled circle with error bar is from data, the pink filled histogram, normalized arbitrarily, stems from simulated WS signal events, and the arrows in the inset figure show the edges of the signal region; (b) RS events from data, where the filled circles with error bars are from data, the blue solid line represents the result of the fit and the dashed line shows the background contribution.}
  \label{fig:lambda_osc}
\end{figure}

\section{Summary}

The BESIII experiment provides a unique laboratory for searches for new physics in the baryon sector. Using large and clean charmonium data samples, BESIII has performed a series of searches for invisible baryon decays, dark baryons, and
baryon-number--violating processes. No significant deviations from the SM are observed in the current data.

The resulting limits place strong constraints on models involving light dark sector particles, baryonic dark matter, axion-like particles, and $\Delta B=2$ interactions. These results demonstrate the power of precision measurements at the intensity frontier and highlight baryons as sensitive probes
of physics beyond the SM. With larger data samples already collected and under analysis, BESIII will continue to push the frontier of baryonic new
physics searches.

The author would like to thank the Baryons 2025 conference organizers for their excellent hospitality and for creating an engaging atmosphere for physics discussions.

This work is supported by National Research Foundation of
Korea under Contract No. NRF-2022R1A2C1092335.







\end{document}